# Highly efficient interaction of a tubular-lattice hollow-core fiber and flexural acoustic waves: design, characterization and analysis

Ricardo E. da Silva, Jonas H. Osório, Gabriel L. Rodrigues, David J. Webb, Frédéric Gérôme, Fetah Benabid, Cristiano M. B. Cordeiro and Marcos A. R. Franco

The modulation efficiency of a tubular-lattice hollow-core fiber (HCF) by means of flexural acoustic waves is investigated in detail for the first time. The main acousto-optic properties of the HCF are evaluated employing 2D and 3D models based on the finite element method. The induced coupling of the fundamental and first higher-order modes is simulated in the wavelength range from 743 to 1355 nm. Significant acoustic (amplitude, period, strain, energy) and optical parameters (effective index, beat length, birefringence, coupling coefficient) are analyzed. The simulations are compared to experimental results, indicating higher modulation performance in HCFs compared to standard optical fibers. In addition, useful insights into the design and fabrication of all-fiber acousto-optic devices based on HCFs are provided, enabling potential application in tunable spectral filters and mode-locked fiber lasers.

## I. INTRODUCTION

ACOUSTO-OPTIC modulation of optical fibers has attracted considerable attention in the last years, enabling remarkable applications in tunable spectral filters, fiber sensors, Q-switched and mode-locked fiber lasers [1], [2], [3], [4], [5]. The all-fiber acousto-optic modulators (AOMs) usually provide electrical tuning and easy integration with the current fiber optic components and devices. Additionally, AOMs can be fabricated employing a few components, such as a piezoelectric transducer (PZT), an acoustic horn, and a segment of optical fiber. In general, the fiber's spectral and power properties are tuned by the frequency and amplitude of the electrical signal applied to the PZT. Thus, flexural acoustic waves have been extensively investigated to couple the power of propagating optical modes in several kinds of optical fibers, such as single-mode and few-mode fibers [2], [3], fibers to compensate dispersion [6], [7], [8], photonic crystal fibers [9], and hollow-core fibers [1], [10], [11]. The resulting spectral filters usually provide significant broad wavelength tuning ranges (18 – 1000 nm) and bandwidths (1 – 20 nm) [1], [3], [7], [9], [11], employing devices with relatively short switching times (200 µs).

In this context, standing acoustic waves are suitable for mode-locked pulsed fiber lasers [4], [12]. The acoustic waves modulate the amplitude of the transmitted optical power with a repetition rate of twice the acoustic frequency. Besides, the modulated bandwidth defines the number of locked axial modes to generate the optical pulses. Consequently, devices modulating broad spectral bands might effectively use the gain bands of the active media commonly employed in fiber lasers, generating short pulse widths with high peak powers. The use of short fiber lengths in AOMs is therefore suitable to increase the modulated bandwidth, contributing to shortening the laser pulse width [12].

In standard optical fibers, the acoustic energy is mostly distributed over the fiber cladding reducing the overlap with the optical modes in the fiber core. AOMs using fibers with reduced diameters, long fiber lengths, and high voltages have therefore been employed to enhance the acousto-optic interaction [2], [3], [4], [5], [6], [7], [8], [13], [14]. In particular, cladding reduction using etching or tapering techniques increases the acoustic amplitude and field overlap between the coupled modes in the fiber cross-section [13], [14]. However, relevant reduced diameters might change the fiber's original modal properties and transmission spectrum. For example, standard single-mode fibers (SMFs) can become effectively multimode when tapered to selected diameter values [14], which might expose the guided cladding modes to external contaminants on the fiber surface. Furthermore, the etched or tapered fiber becomes fragile and susceptible to unsuitable deformations, macro-bends, mechanical instability, or fracture. Overall, the use of high voltages and long fiber lengths might increase the consumed energy and switching time response of the devices.

We have recently experimentally demonstrated a highly efficient acousto-optic spectral filter based on a tubular-lattice hollow-core fiber (TL-HCF) [1]. These HCFs provide a large protective silica jacket for the guided optical modes, minimizing the effect of the undesired perturbations from the environment. Moreover, HCFs usually offer low losses over the wavelength range of important active media, such as erbium and ytterbium, providing low nonlinearity and high

This work was supported by the grants 2022/10584-9, São Paulo Research Foundation (FAPESP), 310650/2020-8, 309989/2021-3 and 305024/2023-0, Conselho Nacional de Desenvolvimento Científico e Tecnológico (CNPq), and RED-00046-23, Minas Gerais Research Foundation (FAPEMIG).

R. E. da Silva and M. A.R. Franco are with Institute for Advanced Studies (IEAv), São José dos Campos, 12228-001, Brazil.

J. H. Osório, G. L. Rodrigues and C. M. B. Cordeiro are with the Institute of Physics Gleb Wataghin, University of Campinas (UNICAMP), Campinas, 13083-859, Brazil. J. H. Osório is also with the Department of Physics, Federal University of Lavras (UFLA), Lavras, 37200-900, Brazil.

D. J. Webb is with the Aston Institute of Photonic Technologies (AIPT), Aston University, Birmingham, B4 7ET, UK. R. E. da Silva is also a visiting fellow at AIPT, Aston University ( r.da-silva@aston.ac.uk).

F. Gérôme and F. Benabid are with the GPPMM Group, XLIM Institute, UMR CNRS 7252, University of Limoges, Limoges, 87060, France.

damage threshold for applications in high-power fiber amplifiers and lasers [15].

Here, we study in detail how the HCF's geometry improves the modulation efficiency and performance compared to SMFs. Additionally, we demonstrate a step-by-step guideline to design and characterize the coupling strength and spectral tuning response of practical devices based on the 2D/3D finite element method (FEM). This study is organized as follows: Section II introduces the analytical formulation employed to calculate selected acoustic and optical parameters of the HCF. Section III-A describes the modeling of a real HCF of 200 µm diameter (HCF-200) based on FEM. An ideal 125 µm HCF (HCF-125) and an SMF are also simulated for comparison. The results show the effect of the HCF's tubes and air core on the acoustic amplitude, strain, and energy. Section III-B evaluates the influence of the HCF's structure on the optical modal overlap and modulation strength in the HCF-125 and SMF. A study about the acoustically induced birefringence in the HCF-200 is demonstrated in Section III-C, by comparing the simulations with experimental evidence. Overall, the HCF's acousto-optic modulation performance and efficiency are discussed, and the main properties significant for practical fiber-based devices are analyzed.

## II. THEORETICAL BACKGROUND

Flexural acoustic waves induce a longitudinal strain along an optical fiber, changing the refractive index over the fiber cross-section. This index variation is based on two mechanisms of opposite contributions: first, the induced strain alters the refractive index by the elasto-optic effect; second, the fiber's geometric deformation changes the optical path length of the guided modes [13], [16]. Fig. 1(a) illustrates the longitudinal strain in a TL-HCF induced by a flexural acoustic wave of amplitude $A$, period $\Lambda$, and frequency $f$. The positive strain (in red) reduces the material refractive index while increasing the optical path length. The geometric effect effectively dominates, causing a net change in the refractive index according to [16],

$$\Delta n(x,y,z,t) = n_0(1+\chi)S_z(x,y,z,t), \quad (1)$$

which can also be written as,

$$\Delta n(x,y,z,t) = \Delta n(x,y)\cos(\omega t - \Omega z), \quad (2)$$

where,

$$\Delta n(x,y) = n_0(1+\chi)\Omega^2 A y, \quad (3)$$

$S_z$ is the longitudinal strain along the fiber, $n_0$ is the refractive index of the unbent fiber, $\chi = -0.22$ is the elasto-optic coefficient of silica, $\omega = 2\pi f$ is the angular frequency, and $\Omega = 2\pi/\Lambda$ is the acoustic wavenumber. For an acoustic wave vibrating in the $yz$ plane, the modulated refractive index over the fiber cross-section caused by the induced bend is approximated as [17],

$$n(x,y) = \left(n_0^2 + A(1+\chi)\left(\frac{2\pi}{\Lambda}\right)^2 2y\right)^{\frac{1}{2}}. \quad (4)$$

The modulation of the refractive index effectively couples power between the fundamental mode $LP_{01}$ and a higher-order

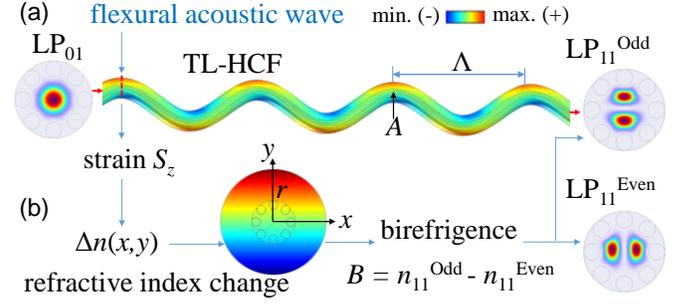

**Fig. 1.** (a) Illustration of a tubular lattice hollow-core optical fiber (TL-HCF) modulated by a flexural acoustic wave of period $\Lambda$ and amplitude $A$. The periodic bends induce a longitudinal strain $S_z$ along the fiber, coupling power between the fundamental mode $LP_{01}$ and the higher-order mode $LP_{11}$ (the power distribution of the modes is shown in the fiber core). (b) The induced non-uniform variation of the refractive index $\Delta n(x,y)$ over the HCF cross section causes modal birefringence splitting $LP_{11}$ in the odd and even modes.

mode $LP_{1m}$ at the resonant wavelength $\lambda_C$, when the optical beat length [3], [6], [10],

$$L_B = \frac{\lambda_C}{n_{01} - n_{1m}}, \quad (5)$$

matches with the acoustic period [16],

$$\Lambda = \left(\frac{\pi r c_{ext}}{f}\right)^{\frac{1}{2}}, \quad (6)$$

where, $r$ is the fiber radius and, $c_{ext} = 5740$ m/s is the silica's extensional acoustic velocity [18]. The phase-matching condition derived from (5) and (6) as, $L_B = \Lambda$, enables the acoustically induced tuning of $\lambda_C$ as,

$$\lambda_C = (n_{01} - n_{1m})\Lambda, \quad (7)$$

where, $n_{01}$ and $n_{1m}$ are respectively the effective refractive indices of the modes $LP_{01}$ and $LP_{1m}$. The modulation depth of the resonance centered at $\lambda_C$ depends on the coupling coefficient as [16],

$$k_C = \frac{\pi}{\lambda}\int_{Area}\psi_{01}(x,y)\,\Delta n(x,y)\psi_{1m}(x,y)dxdy, \quad (8)$$

where, $\psi_{01}$ and $\psi_{1m}$, are the normalized electric field distribution of the modes $LP_{01}$ and $LP_{1m}$ over the fiber cross-section. Thus, the modulation depth and resonant wavelength $\lambda_C$ are properly tuned by changing the acoustic amplitude $A$ in (3) and frequency $f$ in (6).

In certain cases, flexural acoustic waves may cause birefringence in the guided optical modes [2], [7]. Fig. 1(b) illustrates the acoustically induced refractive index change, $\Delta n(x,y)$, over an HCF's cross section of radius $r$ ($\Delta n$ shares the same color scale of $S_z$). Note that the radial variation of $\Delta n(y)$ in the $y$-direction differs than $\Delta n(x)$ in the $x$-direction. The resulting modal birefringence, $B = n_{11}^{Odd} - n_{11}^{Even}$, splits the effective index of $LP_{11}$ in the modes $LP_{11}^{Odd}$ and $LP_{11}^{Even}$, as illustrated in Fig. 1. Previous studies in other fibers have shown that this modal birefringence follows a complex function of the acoustic amplitude $A$, which might be tuned to modulate only one or both even and odd $LP_{11}$ modes [2].

The integrated time-averaged kinetic energy density over the fiber cross-section depends on the displacement components $u$ along the fiber [19],

$$E_k = 2\pi^2 \rho c_t \left(\frac{fD}{c_t}\right)^2 \int_0^{2\pi}\int_0^1 \left(|u_r|^2 + |u_\varphi|^2 + |u_z|^2\right) r_n d\varphi dr_n, \quad (9)$$

where, $D$ is the fiber diameter, $c_t$ = 3764 m/s is the transversal acoustic velocity and, $\rho$ is the silica density. $r_n = r/D$ and $\varphi$, indicate, respectively, the energy variation along the radial $r$ and azimuthal $\varphi$ coordinates in the fiber cross-section.

The acoustic and optical parameters described in (1)-(9) provide a useful tool to characterize the spectral tuning response and modulation efficiency of an optical fiber. A selection of these parameters has been numerically computed for the HCFs and SMF evaluated in this study. The modeling method, results and discussion are described in the next sections.

### III. RESULTS AND DISCUSSION

*A. Influence of the HCF's Geometry on the Acoustic Performance and Efficiency*

We have modeled the cross-section of the actual HCF-200 as shown in Fig. 2(a). The fiber is composed of 8 tubes of 10.7 µm diameter and 300 nm thickness, forming a 30 µm diameter air core [1]. The HCF is modeled employing the package COMSOL Multiphysics based on the FEM. The fiber cross-section is designed in the $xy$ plane, as illustrated in Fig. 2(b). The resulting 2D component is extruded, generating a 7.7 cm long 3D geometry (Fig. 2(c)). The fiber's material parameters are namely the silica density $\rho$ = 2200 kg/m$^3$, Young's modulus $Y$ = 72.5 GPa, and Poisson's ratio $v$ = 0.17 [18]. A sinusoidal force with constant amplitude of $F = 1.11\times10^{-1}$ N is transversally applied at the fiber end ($y$-direction) from $f$ = 190 to 580 kHz (10 kHz step). $F$ induces an acoustic amplitude of about $A$ = 5 µm as estimated for the experimental resonance at $f$ = 196.7 kHz [1]. The other fiber end is fixed. This setup mimics the transversal excitation by the acoustic horn employed in the experiments [1]. Additionally, the HCF-125 and the SMF are modeled by using the same methods and parameters to compare the effect of the HCF's geometry on the acousto-optic modulation efficiency and performance (considering, for the SMF, an 8.2 µm core diameter, 0.36% core-cladding index contrast). The displacements, longitudinal strain and kinetic energy are computed along the fibers for the considered frequency range.

Fig. 2(d) shows an arbitrary flexural acoustic wave in the HCF-200 indicating the maxima and minima (nodes) displacements. Note that the displacements are uniform over the fiber cross-section at every position along the fiber length. The acoustic peak amplitude $A$ and period $\Lambda$ are estimated from the displacement component in the $y$-direction. In turn, Fig. 2(e) shows the longitudinal strain $S_z$ in the fiber cross-section and along the length due to the bending induced by the acoustic modulation (red – positive strain, blue – negative strain). In addition, the kinetic energy is integrated along the whole 3D fiber and in the tubes to estimate the effective energy fraction consumed by the optical modulation. The $S_z$ modulus is further averaged along the HCF tubes at each frequency step. The HCF's acoustic resonances change the strain $S_z$ with frequency, consequently changing the refractive index. Indeed, the silica tubes are the only region of the HCF microstructure that effectively modulates the optical modes in the air core.

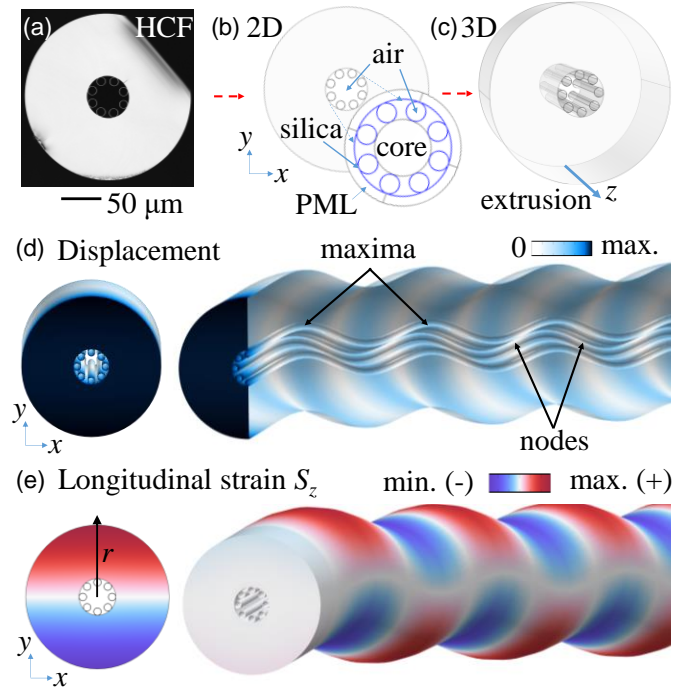

**Fig. 2.** Modeling the acousto-optic properties of the HCF with the finite element method (FEM): (a) The HCF cross-section is modeled as a (b) 2D geometry setting the silica and air regions. The inset in (b) shows the perfectly matched layer (PML) used to reduce the silica domain to compute the optical properties only. (c) The whole 2D geometry is further extruded to generate a 3D fiber model. (d) 3D FEM simulation of a flexural acoustic wave in the HCF indicating the maxima and minima displacements, and (e) the longitudinal strain distribution in the fiber cross-section and along the length.

Fig. 3 shows plots of the 3D flexural acoustic waves relevant parameters corresponding to the HCF-200 (in green), HCF-125 (in red), and the SMF (in blue) from $f$ = 190 to 580 kHz. Fig. 3(a) shows that the HCF-200 resonances have lower acoustic amplitudes compared to the other fibers due to the lager fiber diameter (which might work as a filter at certain frequencies reducing the resonances' number). The SMF has a high amplitude resonance nearly $f$ = 460 kHz. The HCF-125 shows similar resonances due to the same fiber diameter but with acoustic amplitudes up to 38x higher than that of the SMF. This is due to the HCF's air regions that reduce the silica content in the fiber cross-section. The amplitude $A$ averaged over the considered frequency range is shown for both fibers in Table I.

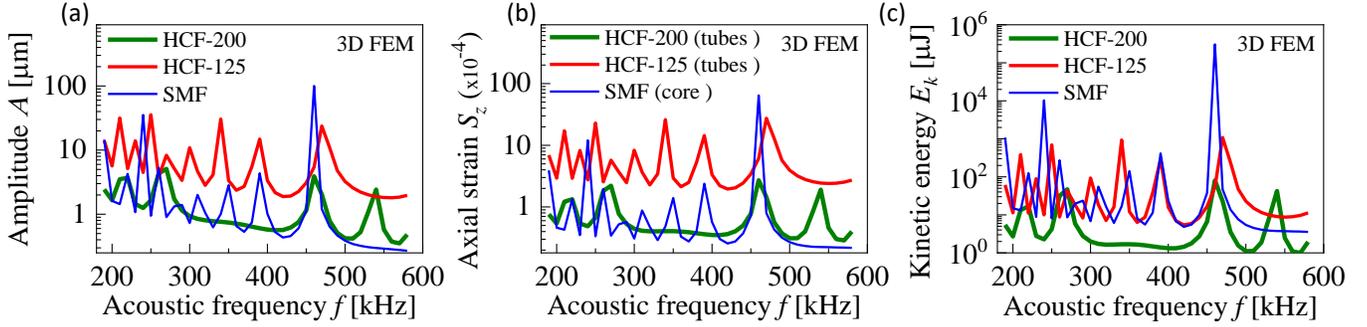

**Fig. 3**. 3D FEM simulation on the acoustic parameters for the HCF-200, HCF-125 and SMF for the frequency range from $f$ = 190 to 580 kHz: (a) acoustic wave peak amplitude $A$, (b) average axial strain along the HCFs' tubes and in the SMF's core and, (c) integrated total kinetic energy $E_k$ along the HCFs and SMF.

TABLE I
3D FEM SIMULATED ACOUSTIC PARAMETERS

| Acoustic parameters | HCF-200 total | HCF-200 tubes | HCF-125 total | HCF-125 tubes | SMF total | SMF core |
|---|---|---|---|---|---|---|
| Amplitude $A$ (μm) | 1.34 | | 7.3 | | 4.8 | |
| Axial strain $S_z$ (μ) | 263 | 70 | 1440 | 604 | 3850 | 250 |
| $S_z$ fraction (%) | - | 27 | - | 42 | - | 6.5 |
| Kinetic energy $E_k$ (μJ) | 8.4 | 0.02 | 108 | 0.8 | 7960 | 34 |
| $E_k$ fraction (%) | - | 0.24 | - | 0.74 | - | 0.004 |

The parameter values were averaged over the considered acoustic frequency range. $A$ is constant over the fiber cross section. $S_z$ is spatially averaged along the whole fiber length (total), HCFs tubes and SMF core. $E_k$ is spatially integrated along the whole fiber length (total), HCFs tubes and SMF core.

Fig. 3(b) shows the spatially averaged longitudinal strain induced in the HCFs' tubes and SMF's core. Table I shows the corresponding values averaged over the considered frequency range. The strain in HCF-125 tubes is about 9x higher than that in HCF-200 and 2.5x larger than in the SMF core. Table I also displays the strain over the full fiber cross-section (total) and the fraction of the total strain in the HCFs' tubes and in the SMF core. We note that the strain fraction in the HCFs' tubes is up to 6.5x larger than in the SMF's core. This is because the strain increases from zero at the cross-section center to a maximum at the fiber surface – see Fig. 2(e). Consequently, the tubes forming the large HCF core located far from the center are under larger strains compared to the corresponding value in the small SMF core. It contributes to an increased modulation of the refractive index in the HCFs with respect to the SMF.

The total kinetic energy integrated along the fibers is shown in Fig. 3(c). We note that the HCF-200 carries lower average acoustic energy compared to the other fibers due to the smaller displacements along the fiber (Fig. 3(a)). Additionally, the kinetic energy values corresponding to the HCF-125 are significantly 74x lower than those related to the SMF, due to the absence of silica in the air regions. The energy fraction in the HCF-125 tubes is 185x larger than in the SMF core, suggesting that the reduced HCF's diameter contributes to concentrating the input energy in the tubes, the effective silica region for optical modulation.

*B. Influence of the HCF's Geometry on the Optical Modal Overlap and Acousto-Optic Modulation Strength*

The effect of the HCF-125 structure on the modal properties is evaluated and compared to the SMF (HCF-200 is not considered to neglect the influence of the fiber diameter). The 2D geometry employed to compute the optical properties of $LP_{01}$ and $LP_{11}$ modes is similar as previously seen in Fig. 2(b). The refractive index of the air-filled regions is set to $n_0 = 1$ and that of silica is calculated by the Sellmeier equation. The modes' effective refractive indices, $n_{01}$ and $n_{11}$, and electric field distributions, $\psi_{01}$ and $\psi_{11}$, are computed (the fields $\psi$ are polarized in the y-direction and the total energy density $E_\psi$ is integrated over the fibers cross-sections). The fields $\psi$ are then normalized to have the same energy of $E_{T\psi} = 1$ μJ, as, $\psi_{N1} = \psi(E_{T\psi}/E_\psi)^{1/2}$, and additionally normalized to the norm, $\psi_{N2} = \psi_{N1}/(\int \psi_{N1}.\psi_{N1}^*)^{1/2}$, as described in [17]. The acoustically modulated index change $\Delta n(x,y)$ in (3) is computed considering the acoustic amplitude $A = 1$ μm and period $\Lambda = 2.4$ mm for both fibers. The coupling coefficient $k_C$ is further computed using (8). The spectral-frequency response of the resonances is calculated using the computed effective indices, beatlengths and acoustic periods in (5), (6) and (7).

Fig. 4 evaluates the modal overlap, coupling coefficient $k_C$ and spectral tuning of the HCF-125 and SMF at $\lambda_C$ = 850, 1060, 1310, and 1550 nm. Note that $k_C$ in (8) depends on the field overlap between the coupled optical modes and the acoustically modulated change of the refractive index $\Delta n(x,y)$. Fig. 4(a) shows the modulus of the electric fields $|\psi|$ of the modes $LP_{01}$ and $LP_{11}$ calculated along the SMF radius $r$ in the positive y-direction. The modal fields and $\Delta n(r)$ are normalized to their maxima values. $\Delta n(r)$ is modulated by the same acoustic wave for both SMF and HCF. Note in Fig. 4(a) that the SMF's modal fields highly overlap in the fiber core, decaying abruptly with $r$ in the cladding ($LP_{01}$ field decays nearly to zero nulling the overlap). In contrast, the modulated index $\Delta n(r)$ is highly concentrated in the fiber cladding achieving a maximum at the outer surface. Although it is not shown here, the SMS supports also other higher-order modes at wavelengths lower than 1260 nm [20], which can also be evaluated employing the proposed numeric method. $LP_{11}$ becomes a cladding mode with increasing $\lambda_C$ significantly reducing the modal overlap and the coupling coefficient $k_C$ at $\lambda_C$ = 1550 nm, as shown in Fig. 4(c). Alternatively, the HCF provides higher modes' confinement in the fiber core, keeping a stronger field overlap for the considered spectral range. In addition, the large air core increases the modal overlap with the modulated $\Delta n(r)$ (considering mainly the change of optical path length due to the fiber curvature), as shown in Fig. 4(b). Consequently, the HCF provides a higher coupling coefficient $k_C$ compared to the SMF (Fig. 4(c)). The contribution of the modal overlap is analogous to the HCF-200 as the reduction of the silica jacket does not relevantly change the fields of the core modes, even for diameters smaller than 125 μm.

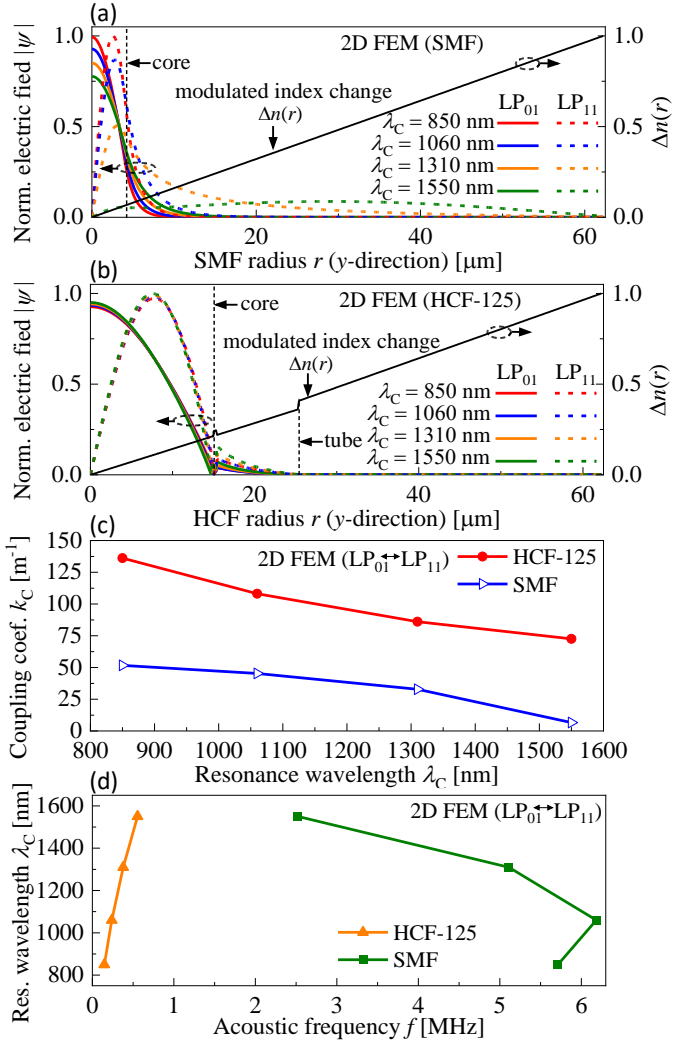

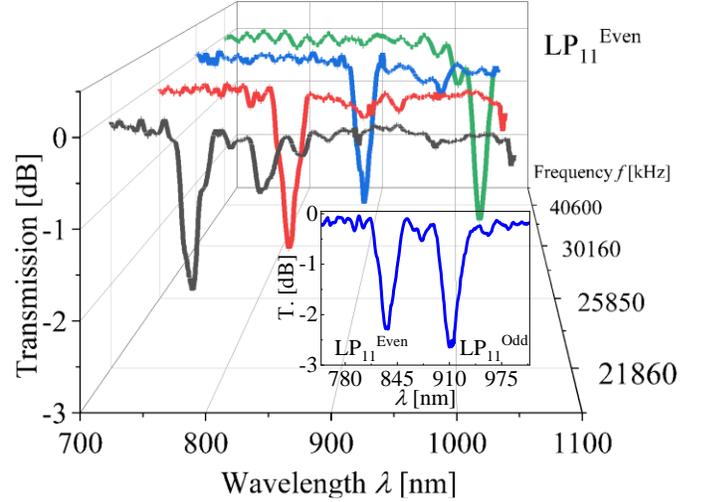

**Fig. 4.** 2D FEM simulation of the radial distribution of the modulus of the electric field $|\psi|$ of the fundamental mode $LP_{01}$ and higher-order mode $LP_{11}$ and radial distribution of refractive index change $\Delta n(r)$ for the (a) SMF and (b) HCF-125 at $\lambda_C$ = 850, 1060, 1310 and 1550 nm. The modal fields are compared to the acoustically modulated $\Delta n(r)$ in the positive y-direction. (c) Corresponding coupling coefficient $k_C$ and (d) spectral tuning with frequency for the HCF-125 and SMF.

The spectral tuning of the considered $\lambda_C$ range for the HCF-125 and SMF is shown in Fig. 4(d). As expected, the frequency response of the SMF is not linear for the overall spectrum. This is because the SMF supports modes with a non-monotonic variation of the modal beatlength in the studied wavelength range [9], [13]. Thus, coupling to other higher-order modes is also expected [4], [13], [14]. In contrast, the HCF provides an almost linear response favoring coupling to the $LP_{11}$ mode in the HCF's large transmission window (the phase-matching to other higher-order modes is not satisfied because of the larger difference in their effective refractive indices). In addition, the tuning slope of 1700 nm/MHz in the HCF is significantly higher than for the SMF (-126 nm/MHz for the descending curve in Fig. 4(d)).

**Fig. 5.** Measured acoustically modulated spectrum of the HCF-200 indicating an example of the spectral tuning of the first resonance (mode $LP_{11}^{Even}$) with increasing acoustic frequency. The inset shows the two modulated resonances caused by the induced birefringence in the fiber cross section. The notches correspond to the even and odd $LP_{11}$ modes.

*C. Study of the Acoustically Induced Birefringence on the Spectral-Frequency Tuning of the HCF*

The HCF-200 is modeled by using the 2D geometry, method and materials previously described in Section III-B. In addition, a perfectly matched layer (PML) of 5 μm in thickness is set around the fiber tubes to provide an electromagnetic absorbing boundary condition [21], as shown in the inset in Fig. 2(b). The PML prevents reflections of the modes' fields at the ending silica interface. It reduces the model's size, decreasing the computing requirements and solution time. The modes' effective refractive indices, $n_{01}$ and $n_{11}$, are computed in the wavelength range of 743 - 1355 nm. The acoustically induced birefringence is further investigated by computing the bend-induced changes in the refractive index over the HCF cross-section using (4) (the elasto-optic contribution is neglected, $\chi = 0$, as the considered optical modes propagate mostly in the air). The input parameters for the simulations are the measured resonances' frequencies and center wavelengths $\lambda_C$ described in [1]. The acoustic amplitude of $A_0 = 5$ μm is estimated from the 3D simulations and measured resonance at $f = 196.7$ kHz, which is nearly the second strongest PZT's resonance as described in [18].

The birefringence in the HCF is investigated in two cases: first, the acoustic amplitude is kept constant at $A = A_0$ for the considered frequency and wavelength range; second, $A$ decays exponentially according to $A = A_0 e^{-fn}$ ($A$ is maximum at the frequency index $f_n = 0$). The spectral-frequency tuning response is calculated employing the modes' effective indices and acoustic periods computed respectively from the 2D and 3D simulations. The numerical curves are further compared to the measured results in [1].

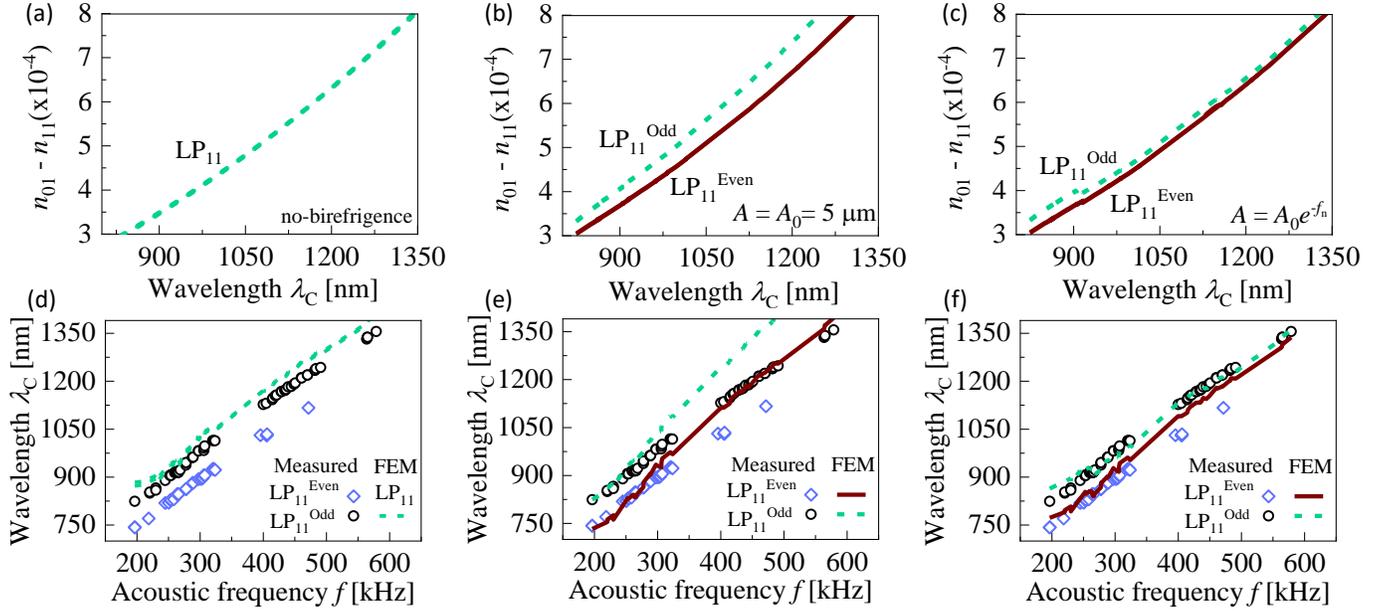

**Fig. 6**. 2D/3D FEM simulation of the induced birefringence caused by flexural acoustic waves in the HCF-200 for the frequency range from $f$ = 196 to 579 kHz for the cases: (a) no-birefringence, (b) constant acoustic amplitude $A_0$ and (c) exponentially decaying amplitude $A = A_0 e^{-fn}$. (d)-(f) show the wavelength $\lambda_C$ tuning of the resonances with increasing frequency corresponding to the cases in (a)-(c). The simulations are compared with the experimental results in [1].

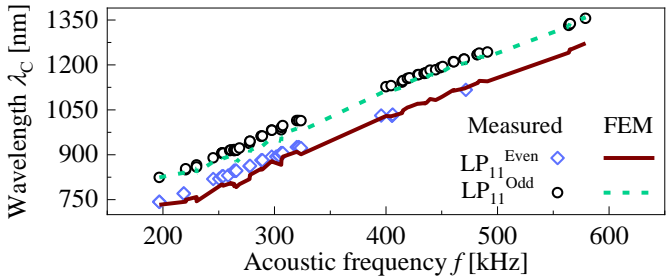

**Fig. 7.** 2D/3D FEM simulation of the spectral-tuning response of the HCF-200 for the frequency range from $f$ = 196 to 579 kHz. The simulated curves are the average of the curves in Fig. 6(e) and (f). The simulations are compared with the experimental results in [1].

Fig. 5 shows the measured modulated spectrum of the HCF-200, indicating the spectral tuning of the mode $LP_{11}^{even}$ with increasing frequency. The inset shows two resonances caused by the acoustically induced birefringence. The notches correspond to the even and odd $LP_{11}$ modes. Fig. 6(a)-(c) show the effective refractive index difference of $LP_{01}$ and $LP_{11}$ for the following cases: (a) no-birefringence, (b) constant acoustic amplitude $A_0$ and, (c) exponentially decaying amplitude $A = A_0 e^{-fn}$. Fig. 6(d)-(f) show a comparison between the simulated and measured spectral tuning responses. Fig. 6(d) indicates ideally negligible birefringence, resulting in only one resonance in the transmission spectrum. This case was employed to estimate the spectral tuning of HCFs [10], [22], however, it cannot predict the induced birefringence as indicated by the measured values in Fig. 6(d). Overall, this case can still be useful to predict one resonance with good agreement between the simulated-measured results (96 %).

The curves' differences in Fig. 6(d) might be caused by minor variations in the HCF geometry, material, and modeling design that could not be predicted in the ideal 2D and 3D models. This difference is therefore neglected in the further two study cases which are intended to fit both curves at the first resonance at $f$ = 196.7 kHz. Thus, only the effects of the acoustic amplitude $A$ in the refractive index are evaluated. Fig. 6(b) shows the modes' index variation for $A_0$ = 5 µm, which is about the average amplitude of the peaks of HCF-200 in Fig. 3(a). It simulates the case in which $A$ does not considerably change at the acoustic resonances' peaks. The simulated-measured values nearly agree up to 350 kHz, deviating their slopes at higher frequencies. Overall, the birefringence gradually increases due to the increasing refractive index with acoustic frequency (reduced period in (4)).

The PZT's effect is considered in Fig. 6(c) and 6(f). The PZT's displacement amplitude is usually higher at the first resonance, reducing with increasing frequency [18]. This approach provides better agreement of simulated and measured results mainly at frequencies higher than 350 kHz, as seen in Fig. 6(f). Hence, the decreasing birefringence reduces the overall tuning range of the $LP_{11}^{Even}$ resonance. In addition, the decreasing $A$ also limits the modulator's maximum tuning range.

Fig. 7 shows the average of the simulated values in Fig. 6(e) and 6(f), indicating that both cases oppositely contribute to an almost constant birefringence, which is observed in most practical devices [10], [22]. The estimated average birefringence is $3.7 \times 10^{-5}$ inducing a 79 nm resonances' separation. Overall, the study of the acoustic and optical properties suggests that the HCF's geometry contributes to improving the acousto-optic modulation performance and efficiency.

## IV. Conclusion

We have proposed a consistent methodology to numerically investigate the main factors that contribute to increasing the acousto-optic modulation performance and efficiency in TL-HCFs. For the best of our knowledge, this is the first detailed study of acousto-optics in TL-HCFs.

For HCF and SMF with the same diameter (125 µm), the HCF's acoustic amplitudes can achieve values up to 38x higher compared to the SMF for the same excitation source. Considering the effective silica regions modulating the optical modes, the average strain induced in the HCF's tubes is about 2.5x larger than in the SMF's core. The strain fraction in the HCF's tubes is 6.5x higher, employing significantly 74x lower total average acoustic energy compared to the SMF. This is because the thin cladding tubes induce higher deformation at regions far from the fiber axis. In addition, the silica reduction in the HCF's tubes and core reinforces the strain in the tubes. Consequently, the deformed tubes significantly change the optical path length of the guided optical modes in the air core. Moreover, the HCF provides a stronger overlap of the coupled optical modes with the acoustic wave over a broad wavelength range ($\lambda_C$ = 850 - 1550 nm), offering an almost linear spectral tuning response. Additionally, the birefringence acoustically induced in the HCF is numerically demonstrated. The good agreement between the simulated and measured results indicates a promising method to characterize also other complex nonuniform mechanical perturbations over the fiber cross section, which might change the properties of the guided optical modes.

These acoustic and optical contributions indicate higher modulation performance and efficiency in TL-HCFs compared to the standard solid fibers. Thus, HCFs with reduced diameters and large cores are promising to further improve the modulation efficiency. In addition, the demonstrated modeling methods might be useful to model other specialty optical fibers, such as, photonic crystal fibers, microstructure optical fibers, or any other fiber with unusual geometries, in which analytical solutions are unavailable or difficult to implement. Thus, the FEM modeling might be applied to characterize the sensitivity and efficiency of several fiber-based devices, such as acoustic and strain fiber sensors. Further advance might come from investigation of HCFs using other dimensions and designs. In this framework, the flexibility to change the HCF geometry might be useful to adjust the modulated optical spectrum, bandwidth, and modulation depth, hence enabling great potential for fast all-fiber tunable filters, pulsed fiber lasers and fiber sensors.

## Acknowledgment

The authors thank F. Amrani and F. Delahaye with the GPPMM Group, XLIM Institute, University of Limoges, for contributing to the fiber fabrication.